\documentclass[12pt]{iopart}
\usepackage{iopams}  
\usepackage{graphicx}

\usepackage{color}  

\usepackage{cite}
\usepackage{multicol}
\def\eqnref#1{Eq.~(\ref{#1})}
\def\figref#1{Fig.~(\ref{#1})}
\def\appref#1{\ref{#1}} 
\begin{document}

\title{Resonant low-energy electron attachment to O\(_{2}\) impurities in dense neon gas}

\author{A. F. Borghesani}
\address{CNISM Unit, Department of Physics and Astronomy\\ University of Padua, Italy}
\ead{armandofrancesco.borghesani@unipd.it}

\begin{abstract}
We report measurements of resonant low-energy electron attachment to O\(_2\) molecular impurities in neon gas in the temperature range \(46.5\,\mbox{K}\le T\le 101\,\mbox{K}\). The reduced attachment frequency \(\nu_A/N\) shows a well defined peak as a function of the gas density \(N\) when the electron energy is resonant with the 4th vibrational level of O\(_2^-\). For \(46.5\,\mbox{K}\le T\le 48.4\,\)K  a second peak has been detected at a much higher density, which is due to the formation of ions in the 5th vibrational level. The temperature dependence of the first peak position can be explained within an ionic bubble model by computing the electron excess free energy as a function of \(T\) and \(N\). The peak shape is rationalized by taking into account the density dependent shift of the electron energy distribution function and the density of states of excess electrons in a disordered medium, and by assuming that electrons sample the gas density over a region of the order of the ionic bubble radius.
   \end{abstract}

\vspace{2pc}
\noindent{\it Keywords}: resonant electron attachment, dense neon gas, ionic bubble, disordered systems. 
 
\submitto{\PSST}
\maketitle
% \ioptwocol
%%%%%%%%%%%%%%%%%%%%%%%%%
%% main text %%
%%%%%%%%%%%%%%%%%%%%%%%%%
%
\section{Introduction\label{sect:intro}}
Low-energy electron attachment to O\(_{2}\) molecules is a
phenomenon relevant in several processes involving the interaction of radiation with matter~\cite{christophorou1984}. For instance, its important role in the physico-chemical processes occurring in the upper atmosphere~\cite{mason2001,matejcik1997}, in electrical discharges~\cite{Govinda1982,christophorou1984b}, in low temperature plasmas~\cite{stoffels1998}, in the damage of DNA by low-energy electrons~\cite{li2003,oro24463} is well documented. 

The physical attachment process of low-energy (thermal) electrons has been quite extensively investigated, both in dilute gaseous environments~\cite{Pack1966,Herzenberg1969,Shimamori1977} as well as in moderately dense ones~\cite{McCorkle1972,goans1974,christophorou1978}. The electron attachment to the O\(_{2}\) molecule in its vibrational ground state \(v=0\) is  a three-body  process usually described in terms of the so called Bloch-Bradury %(BB) 
two-stage mechanism~\cite{Bloch1935}.
At some specific energy resonant electron capture proceeds via the formation of a temporary negative-ion nuclear-excited Feshbach resonant state~\cite{christophorou1978c,christophorou1984}. 
The kinetic energy of the captured electron
couples with the molecular vibrations and leads to the formation of an ion in a vibrational excited state with quantum number \(v^{\prime}\) according to the reaction   
\begin{equation} 
\mbox{O}_{2}\left(X\,^{3}\Sigma^{-}_{g};v=0\right)+e \rightarrow  \mbox{O}_{2}^{-\star}\left(X\,^{2}\Pi_{g};v^{\prime}\ge 4\right)
\label{eq:BB1}\end{equation}
The metastable ion, which otherwise would rapidly undergo a quick unimolecular decomposition, can be stabilized in a dense gaseous environment by collision with a third body \(M\), typically a host atom that carries away the excess energy~\cite{Illenberger1994}
\begin{equation}
\mbox{O}_{2}^{-\star}\left(X\,^{2}\Pi_{g};v^{\prime}\ge 4\right)+ M  \rightarrow  \mbox{O}_{2}^{-}\left(X\,^{2}\Pi_{g};v^{\prime}< 4\right)+M
\label{eq:BB2}\end{equation}
\begin{figure}[t!]\centering\includegraphics[width=\textwidth]{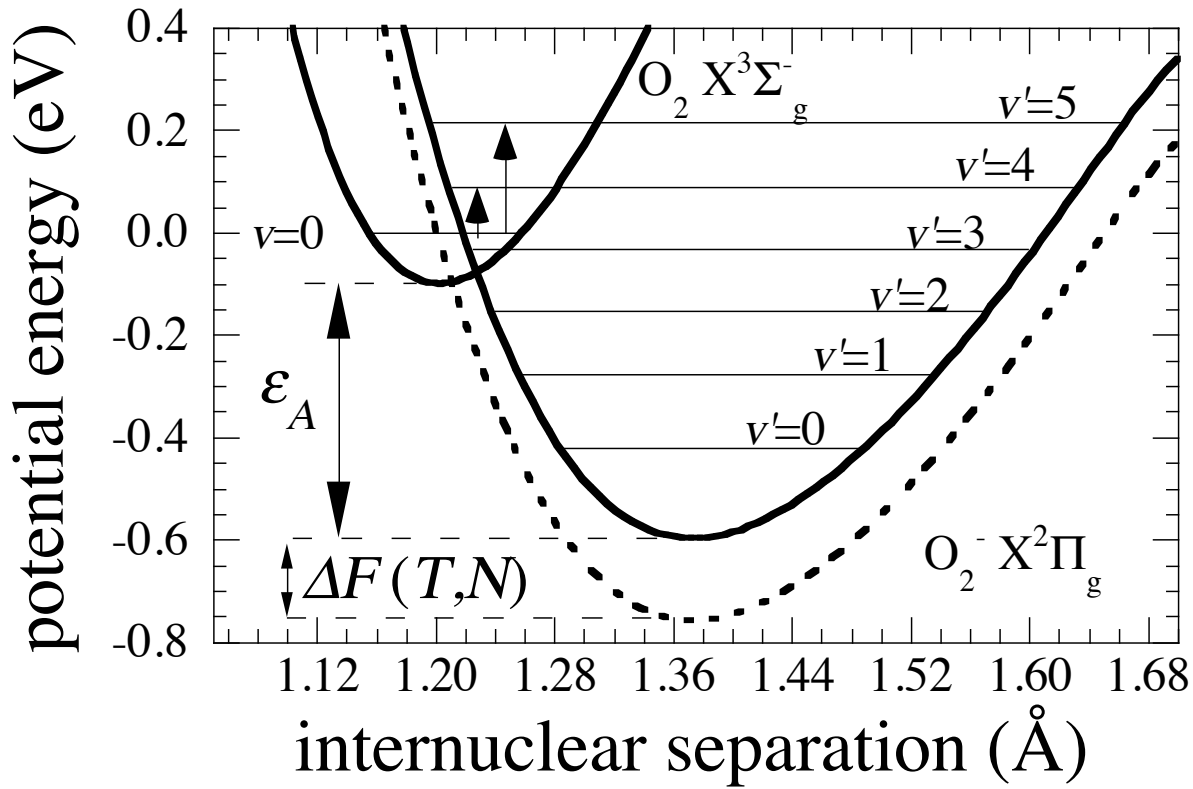}\caption{\small Potential energy curves for O\(_{2}\) and O\(_{2}^-\)~\cite{Schulz1970}. The O\(_2^-\) curve is shifted with \(N\) by \(\Delta F \), the excess free energy (dotted line). At zero density \(-\Delta F = \epsilon_{A}\), where \(\epsilon_{A}\) is the electron affinity. The arrows indicate the transitions to the two first vibrational levels of the ion available for attachment  (see text).
\label{fig:potenziali}}
\end{figure}
The potential energy curves of the neutral molecule and of its anion are shown in ~\figref{fig:potenziali}. The negative-ion state lies energetically below the ground state of the parent molecule and exhibits a positive electron affinity (\( \epsilon_{A}\approx 0.46\,\)eV~\cite{Pack1966,Celotta1972a,Travers1989,Ervin2003}). The electron attachment turns out to be a resonant process because the energy \(\epsilon\) of the colliding electron must equal the energy difference between the energy of the excited ion state \((\epsilon_{v^{\prime}}-\epsilon_{A})\)
and that of the ground state of the neutral molecule \(\epsilon_\mathrm{gs}\),
\(\epsilon=\epsilon_{Rv^{\prime}}\equiv \epsilon_{v^{\prime}}-\epsilon_{A}-\epsilon_\mathrm{gs}\).
The resonances show a doublet structure due to spin-orbit coupling in the molecular ion~\cite{Land1973,Land1974}. The energies of the center of the \(v^{\prime}=4\) and \(v^{\prime}=5\) resonances are approximately located at  \(\epsilon_{R4}=90\,\)meV and \( \epsilon_{R5}=210\,\)meV above the neutral ground state, respectively~\cite{Spence1970,spence1972,Celotta1972a}.  
In \figref{fig:potenziali} the arrows show the energy difference for the two first available ion states with \(v^{\prime}= 4\) and \(v^{\prime}=5.\)

In a dense gas, which is  archetypal to a disordered system, it is known that the average energy of an excess electron is shifted with respect to the thermal energy by the multiple scattering induced, density-dependent energy at the bottom of the conduction band \(V_{0}(N)\)~\cite{Springett1968a}. 
Thus, the resonant character of the attachment process gives the researchers the unique possibility to directly investigate the energetics and statistics of excess electrons in a dense disordered medium, thereby yielding useful pieces of information on the electron density of states (DOS) and energy distribution function. 

The first experimental evidence of resonant electron attachment to O\(_{2}\) in dense helium gas was produced by Bartels~\cite{bartels1973}. At \(T=77\,\)K  a sharp peak
in the attachment frequency \(\nu_{A}\) was observed at a density \(N_{4}\approx 30\times 10^{26}\,\)m\(^{-3}\) as well as the low-density shoulder of a second peak at higher densities. As the thermal energy amounts to only a few meV, the most important contribution to the resonance energy is believed to be supplied by \(V_{0}(N)\), which, at the densities of the experiment, can be expressed by the Fermi shift~\cite{fermi1934}
\begin{equation}
V_{0}(N)= \frac{2\pi\hbar^{2}}{m}Na
\label{eq:Fermishift}\end{equation}
in which \(a\) is the electron-atom scattering length, \(\hbar=h/2\pi\), \(h\) is the Planck's constant, and \(m\) is the electron mass. At the density \(N_{4}\) of the first peak, the Fermi shift exactly equals the  energy for the \(v^{\prime}=4\) resonance, \(V_{0}(N_{4}) = \epsilon_{R4}\).

An unexpected increase of \(N_{4}\) with the temperature \(T\) was later discovered in helium in the range \(50\,\mbox{K} < T< 100\,\mbox{K}\) by using a square-wave technique. The shape of the peak was semiquantitatively reproduced by taking into account both the density dependent energy shift of the excess electrons and the effect of the density fluctuations on the electron DOS~\cite{bruschi1984}. 

Finally, the temperature dependence of \(N_{4}\) in helium was measured in a wider temperature range by using a pulsed Townsend photoinjection technique~\cite{neri1997}. The results confirmed those of the previous experiments. The presence of a peak in the attachment frequency was rationalized as the result of the multiple scattering induced shift of the electron energy distribution function and of the quantum nature of the electron-atom interaction~\cite{Khrapak1995a,volykhin1995,grigorev1999}. 
In this experiment the researchers were also able to observe the almost complete second attachment peak %in helium 
caused by attachment to the \(v^{\prime}=5\) ion state. At the same time it was shown that in argon gas no attachment peaks are present because the resonance condition is never met owing to the much larger atomic polarizability of the argon atoms with respect to helium atoms that leads to \(V_{0}(N) <0\)~\cite{reininger1983}.

 It is to be noted that in all of the mentioned experiments the attachment peak does not show the expected spin-orbit split doublet structure of the ion vibrational levels. This absence has also been noticed in several other crossed beam experiments in which the electron attachment to oxygen clusters formed by nozzle expansion is investigated~\cite{mark1985,mark1986,matejcik1996,mark1999}
 and, up to now, no explanation has ever been given~\cite{Hernandez1991c}. For this reason, in the following the energies at the center of the resonances are taken as the resonance energies. 
 
 Several problems, however, remained unsolved in the helium experiment~\cite{neri1997}. In particular, the shape of the first peak could not easily be reproduced by the use of the shifted thermal distribution function. Actually, the gas is a highly disordered system and electrons sample the density over a small volume, whose size is of the order of the electron thermal wavelength or mean free path. In such small a volume fluctuations strongly influence the electron DOS leading to a broadening of the energy distribution function. Its computation is still a not completely solved issue. The use of a percolation model~\cite{Eggarter1970,Eggarter1971,Eggarter1972} led to an improvement of the simulation of the peak shape though the agreement with the experimental data was not  satisfactory enough. 
 
Aiming at the solution of the several problems raised by the results in helium gas, we have carried out measurements in dense neon gas over a reasonably wide temperature range. Neon has been chosen because the electron-atom is still dominated by the short-range repulsive exchange forces but the polarization interaction with the host atoms is much stronger than in helium. Thus, the electron-atom scattering cross section is much smaller and more rapidly energy dependent than for helium~\cite{omalley1980} and the self energy \(V_{0}(N)\) is still positive though smaller and shows a different density dependence than in helium.

Moreover, the temperature of the investigated isotherms (\(46.5\,\mbox{K}\le T\le 100\,\mbox{K}\)) are much closer to the critical temperature \(T_{c}=44.4\,\)K than it was in the helium experiment. So, fluctuations are expected to be more effective in modifying the electron energy distribution function.

 In this paper we report the experimental results of the electron attachment measurements in neon gas and their rationalization. 

\section{Experimental details\label{sect:expdet}}
 The experimental apparatus and technique have thoroughly been described in literature~\cite{borghesani1988}.
 We only recall here the main features. 
 The experimental cell can be filled with the gas up to a pressure \(P\approx 10\,\)MPa and its temperature is controlled within \(0.01\,\)K. 
 Isolated bunches of typically \(10^{5}\) to \(10^{6}\) electrons of \(\approx 4\mu\,\)s duration are repeatedly injected into the gas by photoelectric effect and drift under the action of a uniform electric field. Owing to attachment, the induced electron current decreases exponentially with time 
  \(I(t)= I_{0}\mathrm{e}^{-\nu_{A}t}\) and 
 is passively integrated in order to improve the signal-to-noise ratio. The numerical analysis of the recorded voltage signal yields the electron and ion drift times, and the attachment frequency $\nu_{A}$~\cite{Borghesani1990e}. 

Molecular oxygen is already present in trace in the gas as an impurity. 
Its concentration is of some parts per million that is sufficient to carry out attachment frequency measurements. Residual impurities such as water vapor, hydrocarbons, and carbon dioxide are removed by recirculating the gas through a liquid N\(_{2}\)-cooled, active-charcoal trap~\cite{torzo1990}. As the O\(_{2}\) impurity content cannot be easily controlled~\cite{bruschi1984}, we adopted the following procedure. For each experimental run, the cell is filled at constant temperature up to the highest pressure with gas of unknown impurity concentration \(C\). The gas is then progressively spilled out of the cell in a stepwise way so that \(C\) remains constant and measurements can be done at the desired \(P\) down to the lowest one. 
Measurements carried out at the same \(T\) in overlapping pressure intervals allowed us to merge data taken in different runs because \(\nu_{A}/N\propto C \). 
 
 The depletion of the electron population because of attachment does not affect its distribution function. At a typical density of \(N\approx 40\times 10^{26}\,\mbox{m}^{-3}\) and with a typical ion concentration \(C\approx 10^{-5}\), the electron mean free path is roughly one order of magnitude smaller than the average ion-ion distance~\cite{Borghesani1992b}. Thus, in a mean electron-oxygen collision time electrons are thermalized by several collisions with the host atoms. Moreover, the attachment does not practically change the O\(_{2}\) number density \(N_\mathrm{O_{2}}=CN\). Actually, in the drift space of volume \(v\approx 5 \times 10^{-6}\,\)m$^{3}$, there are \(\approx 10^{17}\) O\(_{2}\) molecules which are many orders of magnitude more abundant than the injected electrons.
 \section{Experimental results\label{sect:expres}}  
\begin{figure}[t!]\centering\includegraphics[width=\textwidth]{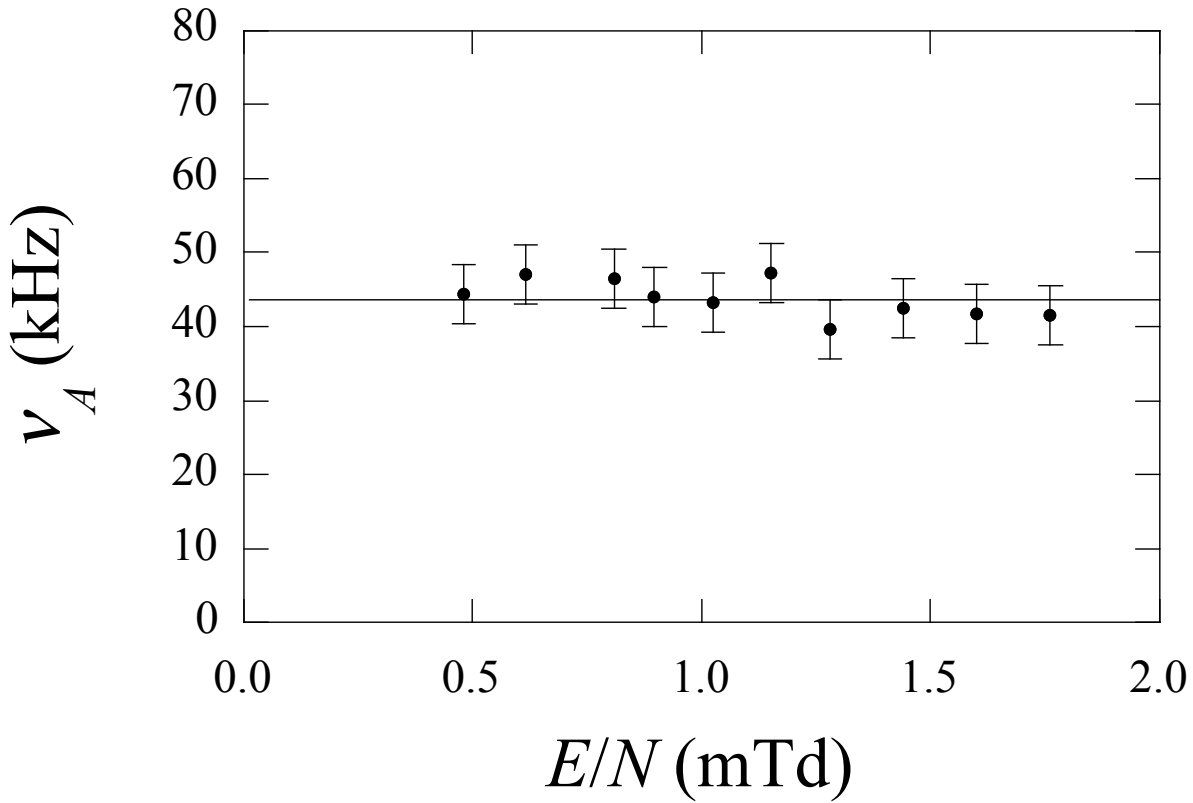}\caption{\small \(\nu_{A}\) vs \(E/N\) for \(N=31.2\times 10^{26}\,\)m\(^{-3}\) at \(T=150.2\,\)K.\label{fig:nuAvsEN}}
\end{figure}

The measurements are carried out in the temperature range \(46.5\,\mbox{K}\le T\lesssim 150\,\)K. The maximum density  attained for each \(T\) is limited by the maximum pressure the cell can withstand.
On each isotherm, \(\nu_{A}\) is measured  on several isopycnals as a function of the applied electric field \(E\) of low strength. The reduced electric field strength \(E/N\) never exceeds \(\approx 5\,\)mTd (\(1\,\mbox{mTd} = 10^{-24}\,\)V\,m\(^{2}\)). The field is so weak as not to significantly heat the electrons and does not modify their distribution function with respect to that at thermal equilibrium. As a consequence, \(\nu_{A}\) does not depend on \(E/N\). A typical example is shown in \figref{fig:nuAvsEN}.

\subsection{Density dependence of the 1st peak  of \(\nu_{A}/N\)\label{sect:nuasunvsn}}
\begin{figure}[t!]\centering\includegraphics[width=\textwidth]{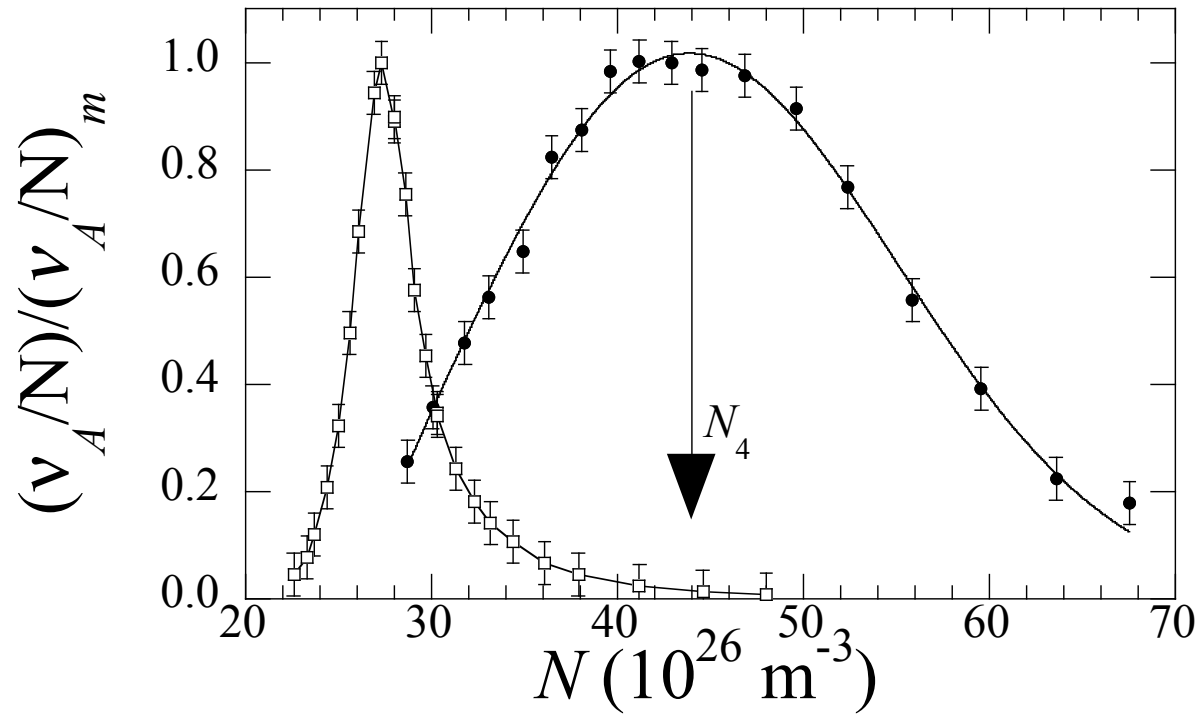}\caption{\small Normalized \(\nu_{A}/N\) vs \(N\) in Ne for \(T=59.8\,\)K (closed points) and in He for \(T=54.5\,\)K (open points)~\cite{neri1997}.  The arrow indicates the density \(N_{4}\) of the maximum.The lines are only a guide for the eye. \label{fig:nuaHe54Ne60}}
\end{figure}

At all temperatures, except the highest at \(T\approx 150\,\)K for which the maximum pressure was not sufficient to reach the necessary density values, \(\nu_{A}\) shows a well defined peak as a function of \(N.\) In order to get rid of the dependence of \(\nu_{A}\) on both the concentration and absolute number of O\(_{2}\) impurities, it is customary to plot the reduced attachment frequency \(\nu_{A}/N\) normalized to unity at the maximum. By so doing, it is possible to directly compare the neon results with those obtained in helium.
In \figref{fig:nuaHe54Ne60} \(\left(\nu_{A}/N\right)/\left(\nu_{A}/N\right)_{m}\) for the neon case is plotted for \(T=59.8\,\)K. For the sake of comparison the data obtained in helium~\cite{neri1997} for the nearby temperature \(T=54.5\,\)K are also plotted.

\begin{figure}[t!]\centering\includegraphics[width=\textwidth]{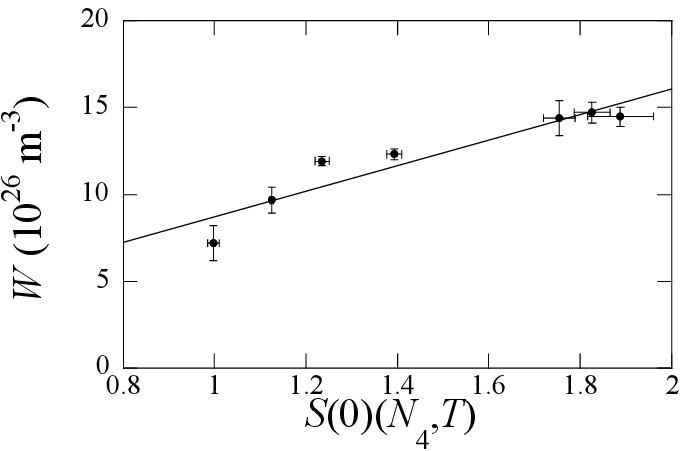}\caption{\small Correlation between 
\(W\) and 
 \(S(0)(N_{4},T)\) in neon gas. The line is a guide for the eye only.\label{fig:WS0}}
\end{figure}

 Two striking differences can be noticed between neon and helium. First of all, the peak density \(N_{4}\) in neon is much higher than in helium. This is not surprising because \(V_{0}(N)\) is much larger in helium~\cite{broomall1976}  than in neon~\cite{borghesani1990} and a higher density value is required in neon to reach the value \(\epsilon_{R4}.\) Secondly, the width of the attachment peak in neon is much larger than in helium at nearly the same temperature.
The peak half width at half height \(W\) is quite well correlated with the long wavelength limit of the static structure factor \(S(0)(N,T)= Nk_\mathrm{B}T\chi_{T}\), where \(\chi_{T}\) is the gas compressibility, evaluated at the density of the peak maximum, as shown in \figref{fig:WS0}.
The neon experiment is carried out much closer to the critical temperature than the helium one. Actually, \(y=\vert T-T_{c}\vert /T_{c}\approx 5\times 10^{-2}\) for neon and \(y\approx 9\) for He.
 We thus draw the conclusion that the density fluctuations  play a much more important role in neon than in helium and will have to be properly accounted for.

 \subsection{The 2nd \(\nu_{A}/N\) peak at high density on  the lowest isotherm\label{sect:2ndpeak}}
\begin{figure}[t!]\centering\includegraphics[width=\textwidth]{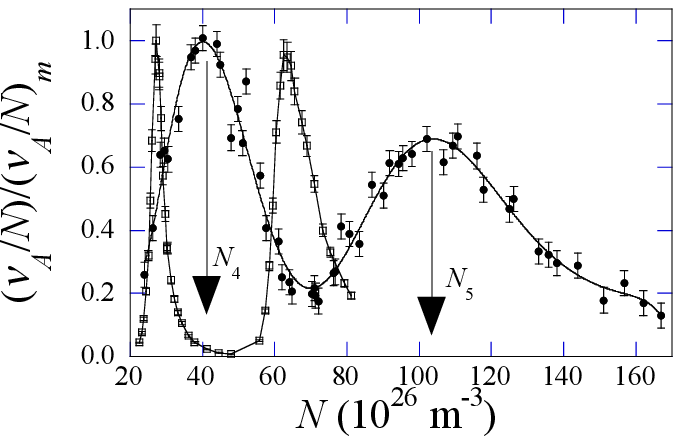}\caption{\small Normalized \(\nu_{A}/N\) in neon for \(T=46.5,\,47.7,\) and \(48.4\,\)K (closed points). For a comparison, the \(T=54.5\,\)K data in helium are also plotted (open points)~\cite{neri1997}. The arrows show the densities \(N_{4}\) and \(N_{5}\) of the two peaks. The lines are a guide to the eye only. \label{fig:2ndpeak}}
 \end{figure}

 On the lowest isotherms \(T=46.5,\,47.7,\) and \(48.4\,\)K, also a second peak at higher densities is present, as shown in \figref{fig:2ndpeak}. This is due to attachment to the \(v^{\prime}=5\) ion level.
 For a comparison also the data obtained in He~\cite{neri1997} are presented. The 2nd peak in neon occurs at the much higher density \(N_{5}\) and is much broader than in helium.  Once more, the phenomenology can be qualitatively explained by both the smaller \(V_{0}\) and by the greater closeness to the critical point of neon. We note that the ratio of the densities of the 2nd to the 1st peaks \(N_{5}/N_{4}= 2.5\) is almost equal to the ratio \(\epsilon_{R5}/\epsilon_{R4}=2.4\) of the resonance energies of the two first accessible ion vibrational levels, as is to be expected if \eqnref{eq:Fermishift} were valid. It is also to be emphasized that the 2nd peak in both gases occurs in a density region in which the phenomenon of electron self-trapping is present~\cite{borghesani1990,borghesani2002}.
  
 \subsection{Temperature dependence of \(N_{4}\)\label{sect:n4t}}
\begin{figure}[t!]\centering\includegraphics[width=\textwidth]{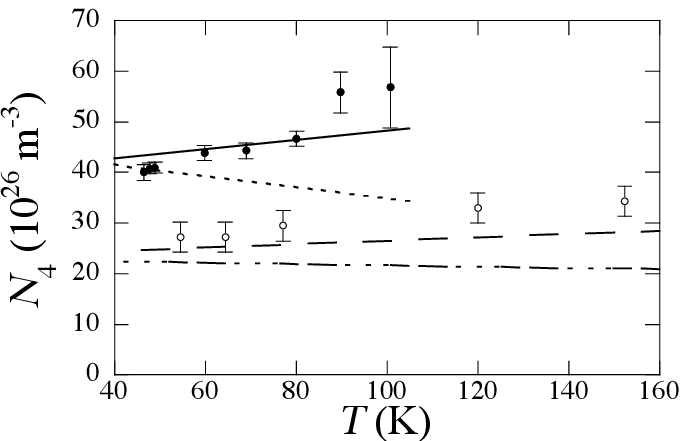}\caption{\small \(N_{4}\) vs \(T\) in neon (closed points) and in helium~\cite{neri1997} (open points). Solid and dashed lines: prediction of the ionic bubble model (\eqnref{eq:rescondDF}, see text). Dotted and dash-dotted lines: energy conservation condition (see text).\label{fig:N4THeNe}}
 \end{figure}

 The density \(N_{4}\) of the 1st peak maximum shows an almost linear, positive dependence on the temperature \(T\), as shown in \figref{fig:N4THeNe}.
This kind of behavior was also detected in helium~\cite{neri1997}. For the sake of comparison the results of helium are also shown in the figure.
The detailed rationalization of the positive slope lines will be given in the discussion of the ionic bubble model~\cite{Khrapak1995a,volykhin1995,grigorev1999}. Suffices it here to say that, on the basis of energy conservation \(\epsilon_{R4}=V_{0}(N)+(3/2)k_\mathrm{B}T\) ($k_\mathrm{B}T/2$ in the case of helium), the  negative slope lines are obtained. 
 % \clearpage
 \section{Discussion\label{sect:disc}}
 The rationalization of the experimental results should  answer the following questions: i) why are there  peaks in the reduced attachment frequency at a specific density, ii) what does determine their shape, and iii) why does the peak density increase with temperature.
\subsection{Relationship between \(\nu_{A}/N\) and the shifted equilibrium distribution function\label{sect:nuaF}} 

Owing to the quite short autoionization lifetime of O\(_{2}^{-}\) in dense gas \(\tau_{a}\approx 2\times 10^{-12}\,\)s~\cite{goans1974}, the capture cross section \(\sigma(\epsilon)\) is strongly peaked at the resonance energy \(\epsilon_{Rv^{\prime}}\) and its width is \(\Delta \epsilon\sim\hbar/\tau_{a}\approx 0.3\,\)meV, to be compared with the width of the electron energy distribution function \(\simeq 4k_\mathrm{B}T\approx 16\,\)meV for \(T=46\,\)K. For this reason the attachment frequency \(\nu_{A}\) can be written as~\cite{bruschi1984}
\begin{equation}
\nu_{A}= p_{s}\sigma_{c}CNw(\epsilon_{Rv^{\prime}})\mathcal{F}(\epsilon_{Rv^{\prime}},T,N)
\label{eq:nuafdist}\end{equation}
in which \(p_{s}\) is the stabilization coefficient, \(\sigma_{c}= \int\sigma(\epsilon)\,\mathrm{d}\epsilon\) is the integrated
capture cross section, and \(C\) is the O\(_{2}\) concentration. The electron velocity \(w\) and the Maxwell-Boltzmann(MB) equilibrium distribution function \(\mathcal{F}\) are to be evaluated at the resonance energy. \(\mathcal{F}\) explicitly depends on the %gas 
density \(N\) because of the shift \(V_{0}(N)\)~\cite{bruschi1984,borghesani1988,borghesani1992}.
The stabilization coefficient is given by\begin{equation}
p_{s}=\frac{k_s N}{\tau_{a}^{-1}+\left(
k_s +k_d\right)N}
\label{eq:ps}\end{equation}
in which \(k_s\) and \(k_d\) are the rates of collisional stabilization and dissociation, respectively.
\(\left(
k_s +k_d \right)\) can be estimated from the measurements of the O\(_{2}^{-}\) ion mobility \(\mu\) in dense neon gas~\cite{borghesani1993,borghesani1995,borghesani2019} as
\(\left(
k_s +k_d\right)= (e/M \mu N ) \) where \(M\) is the atomic mass of neon. At the typical density of the 1st peak, \(N\approx 40\times 10^{26}\,\)m\(^{-3}\) and \(\tau_{a}\left(k_s+k_d\right)N\gtrsim 8\). Thus, \(p_{s}\) can be assumed to be practically density independent. As a consequence, \begin{equation}\frac{\nu_{A}}{N}= D \mathcal{F}(\epsilon_{R},N,T)\label{eq:nuasuNfer}{\tiny }\end{equation} in which \(D\) is a constant.

\begin{figure}[t!]\centering\includegraphics[width=\textwidth]{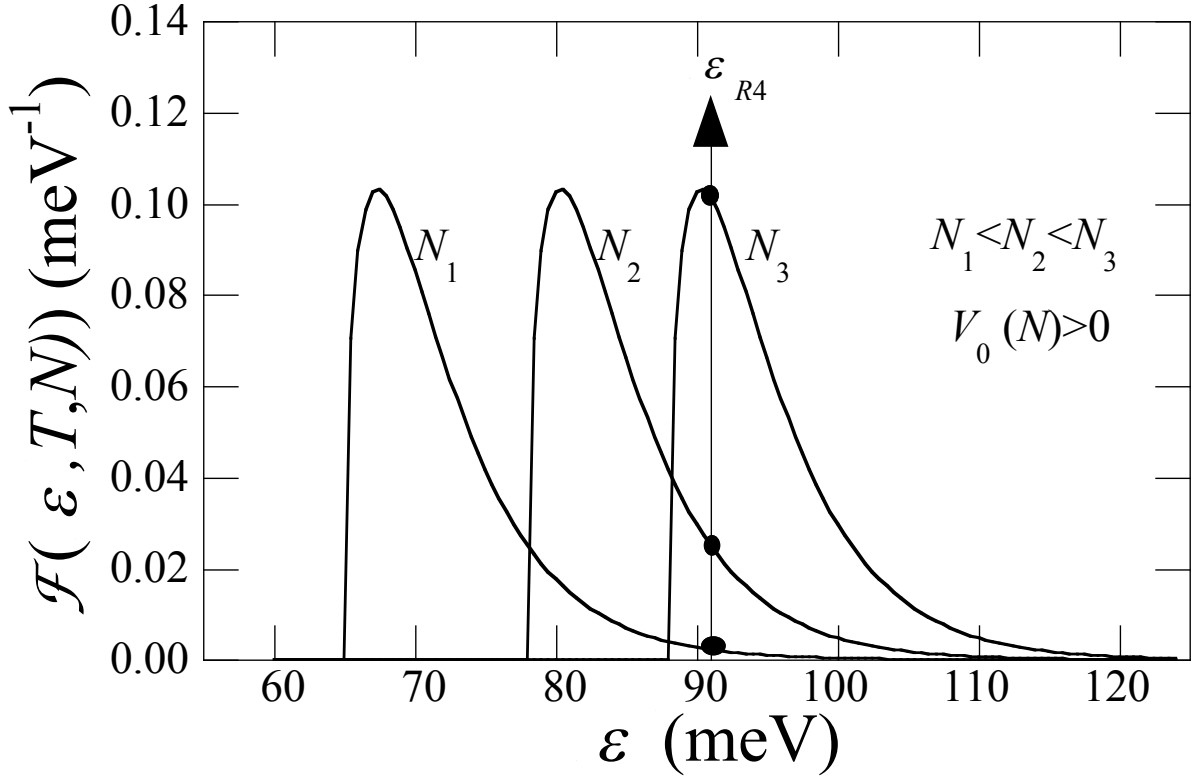}\caption{\small Qualitative picture of the sampling of the shifted MB distribution at constant resonance energy \(\epsilon_{R4}\).\label{fig:shifted}}
\end{figure}

According to the classical kinetic theory~\cite{huxley}, \(\mathcal{F}\) should be independent of density. However, it is now very easy to intuitively grasp the basic features of the attachment peak formation if the density dependent shift of the distribution function is taken for granted. Let us inspect \figref{fig:shifted} and consider the 1st peak.
 The MB equilibrium distribution function is shifted by the amount \(V_0(N)\). The reduced attachment frequency is sampling \(\mathcal{F}\) at the constant energy \(\epsilon_{R4}\).  Thus, by increasing \(N\), at first the low-density shoulder of \(\nu_A/N\) reproduces the high-energy tail of \(\mathcal{F}\). For \(N=N_4\) the maximum of \(\mathcal{F}\) occurs for \(\epsilon_{R4}\) and \(\nu_A/N\) goes through a maximum, as well. Eventually, for higher \(N\), the shift \(V_0\) is such that \(\epsilon_{R4}\) falls in the low-energy tail of \(\mathcal{F}\), which is reflected in the high-density tail of the \(\nu_A/N\).

This na\"ive approach has worked quite well for helium because the temperatures were much higher than the critical one and fluctuations did not play a major role. In neon, however, the situation is more complicated. Actually, if one would simply solve the resonance condition \(\epsilon_{R4}=V_{0}(N_{4})+(3/2)k_\mathrm{B}T\) for \(N_{4}\), one would obtain \(N_{4}\approx 73\times 10^{26}\,\)m\(^{-3}\) for \(T=46.5\,\)K, to be compared with the experimental value \(N_{4}\approx 40\times 10^{26}\,\)m\(^{-3}\). This discrepancy can be traced back to two main reasons. First of all, the quantum nature of the electron-host atom interaction that leads to the formation of an empty cavity around the ion is neglected. The second reason is that fluctuations, so very important in neon, are also neglected. They lead to a DOS broadening that results in a shift of the \(\mathcal{F}\) maximum to lower energies, thereby reducing the required density contribution from \(V_{0}\). 
Evidently, these two mechanisms act simultaneously. We will try to disentangle their effects by first treating the effect of the formation of the void around the ion within the so called {\em ionic bubble} model~\cite{Khrapak1995a,volykhin1995,grigorev1999} and, then, the computation of the distribution function in a disordered medium will be dealt with in the Eggarter's {\em fluctuational model}~\cite{Eggarter1970,Eggarter1971,Eggarter1972}.

\subsection{Predictions of the ionic bubble model\label{sect:ibm}}
The time required for the formation and expansion of a cavity around the ion in neon gas is not known. However, experimental as well as theoretical estimates of the bubble formation time in liquid helium and neon~\cite{hernandez1970,iakubov1982,rosenblit1997} suggest that the formation time is of the order of \(\approx 1\,\)ps, comparable to or even shorter than the autoionization lifetime \(\tau_{a}\). Thus, the cavity around the ion can form before autoionization occurs.

Owing to its relatively small binding energy, the weakly bound outer electron in the ion is localized in a region larger than the electronic shells of the molecule and can be treated as if it were a quasifree electron strongly interacting with the host atoms. As a result of the competition between the short-range exchange repulsive interaction and the long-range attractive polarization one an empty cavity is formed around the ion. A denser layer outside of the cavity  is also formed because of electrostriction~\cite{atkins1959} but, as the neon polarizability is relatively small, it will be neglected  for the sake of simplicity. 
Due to the presence of \(V_{0}(N)\), the binding energy is changed in a cavity whose radius is comparable to the spatial extent of the electron wave function. The optimum cavity radius \(R_{b}\) is determined by minimizing the excess free energy \(\Delta F\), i.e., by optimizing the difference between the increase of the binding energy of the electron in the ion and the energy spent to expand the cavity itself. The details of the computation and minimization of \(\Delta F\) are described in \appref{sect:deltF}.

Once the excess free energy has been minimized with respect to the cavity radius for all densities and temperatures yielding \(\Delta F_{m}\equiv \Delta F (R_{b}, T,N)\), the resonance condition changes to 
\begin{equation}
\epsilon_{Rv^{\prime}}+\Delta F_{m}= V_{0}(N_{v^{\prime}})+\frac{3}{2}k_\mathrm{B}T
\label{eq:rescondDF}\end{equation}
In \figref{fig:N4THeNe} the solid line shows how \eqnref{eq:rescondDF} predicts the change of \(N_{4}\) as a function of \(T\) whereas the dotted line is obtained by only enforcing energy conservation. The agreement with the data is rather good, although it is somewhat worse at higher \(T.\) It appears that the ionic bubble model is also valid in the helium case, for which it produces an even better agreement with the data than in neon, as shown in \figref{fig:N4THeNe}. It is worth emphasizing here that the position of the peak maximum is due to the combined effect of the density dependent shift of the electron energy at the bottom of the conduction band and of the fluctuations induced broadening of the electron energy distribution function. In helium the latter phenomenon is less important because of the distance from the critical temperature and can be neglected. In neon, on the contrary, owing to the greater closeness to \(T_{c},\) the effect of fluctuations contributes a big deal to the location of the maximum and the ionic bubble model can only be used to predict most of the temperature dependence of the maximum location.

\subsection{Shape of the attachment peak\label{sect:peakshape}}
Whereas the concept of the density dependent shift of the distribution function qualitatively rationalizes the presence of a peak in the reduced attachment frequency \(\nu_{A}/N\) and the results of the ionic bubble model rather well explain the increase of the density of the peak maximum with \(T\), nonetheless the explanation of the peak shape  
requires a detailed computation of the distribution function in a disordered medium. To this goal we have adopted the semiclassical fluctuational model~\cite{Eggarter1970,Eggarter1971,Eggarter1972}. A brief description of it and some of its most relevant features will be described in \appref{app:egg}. Suffices here to say that  the density fluctuations lead to the appearance of a low-energy tail of the DOS, whose extension is controlled by the characteristic length \(L\) with which electrons are sampling the distribution of the host atoms.
\(L\) is the most important quantity in the fluctuational model. The original suggestion to describe the electron mobility in dense helium gas at low temperature was to use a length proportional to the uncertainty in the electron position according to the Heisenberg's principle \(L \propto h/\sqrt{2m\epsilon}\), whereas a better agreement for the electron mobility in neon gas was given by choosing a length proportional to the electron mean free path% \(L\propto 1/N\sigma_\mathrm{mt}(\epsilon)\)
~\cite{Borghesani1992b}. 

For the present attachment experiment, in which the attaching electrons have to be somehow localized in a restricted region near the O\(_{2}\) molecule, we have decided to use their thermal wavelength as sampling length
\begin{equation}
L = c(T)\frac{h}{\sqrt{2\pi m k_\mathrm{B}T}}
\label{eq:lt}\end{equation}
in which \(c\) is a temperature dependent adjustable constant.

\begin{figure}[b!]\centering\includegraphics[width=\textwidth]{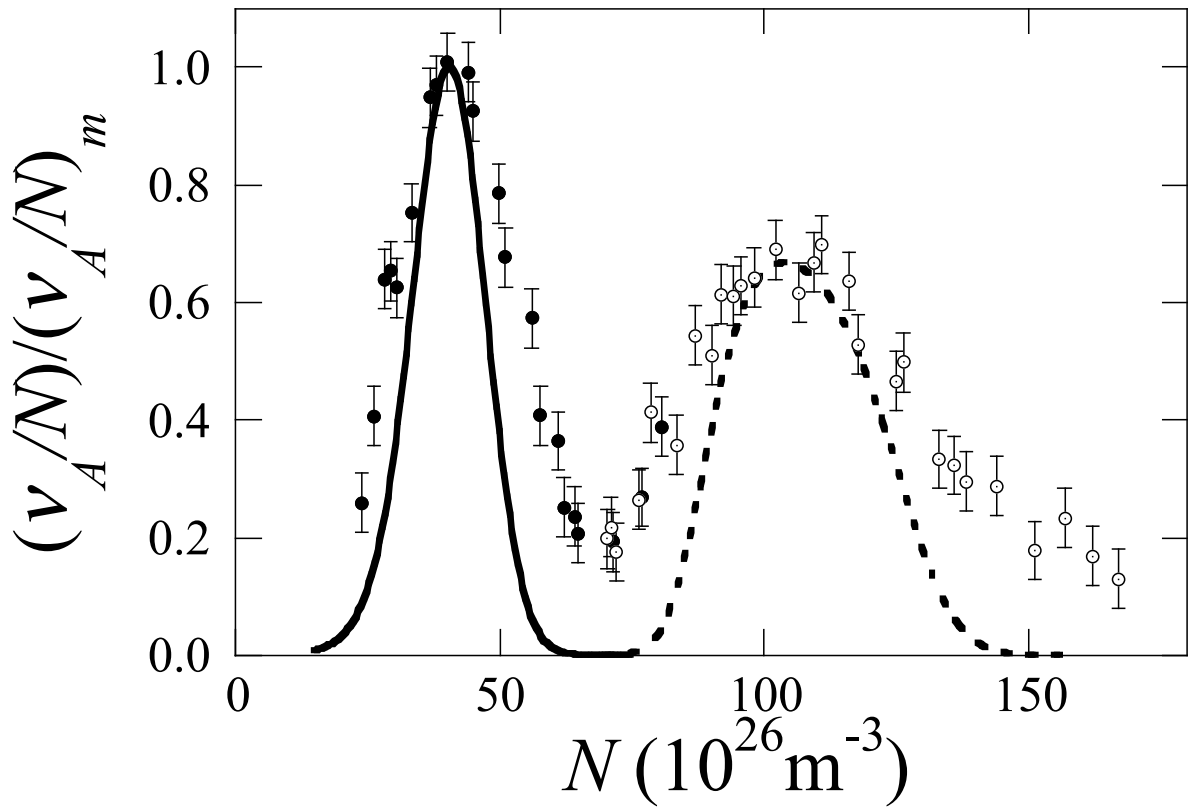}\caption{\small  Normalized \(\nu_{A}/N\) vs \(N\) for the lowest isotherms. \(T=47.7\) and \(48.4\,\)K (closed symbols). \(T=46.5\,\)K (open symbols). Solid line: \(c=0.145\). Dashed line: c=0.353.
\label{fig:NeT47482c0145ec0353}}
\end{figure}

The distribution function, whose shape depends on the choice of \(c\) as shown in \appref{app:egg}, is evaluated at the energy \(\epsilon_{Rv^{\prime}}\) matching the resonance condition \eqnref{eq:rescondDF}. 
A proper choice of \(c\) on the several isotherms leads to satisfactory agreement of the model with the experimental data. In \figref{fig:NeT47482c0145ec0353} we plot the data of the lowest isotherm about \(\approx 47\,\)K.
We note that the two peaks for \(N_{4}\) and \(N_{5}\) are reproduced quite well with different \(c\) values. As the data are measured on an isotherm, the sampling length is larger when \(c\) is larger. This means that the 2nd peak is reproduced only if a larger \(L\) is used than for the 1st peak. A possible rationalization of the meaning of this difference  will be described later. We now note that previous electron mobility measurements have shown that electrons localized in bubbles appear in significant proportion for \(N\ge 95\times 10^{26}\,\)m\(^{-3}\)~\cite{borghesani1990}. This means that the whole high-density tail of the 2nd attachment peak occurs in a region in which most electrons are localized in bubbles. According to literature~\cite{jahnke1975} attachment from self-trapped states should be more efficient than from delocalized states as a probable consequence of the quick stabilization of the ion by the collapsing bubble~\cite{Hernandez1991c}.
This could be the reason why \(\nu_{A}/N\) does not rapidly fall to zero at very high density. In any case the fluctuational model and the ionic bubble one are rather successful at describing also the second attachment peak.

\begin{figure}[b!]\centering\includegraphics[width=\textwidth]{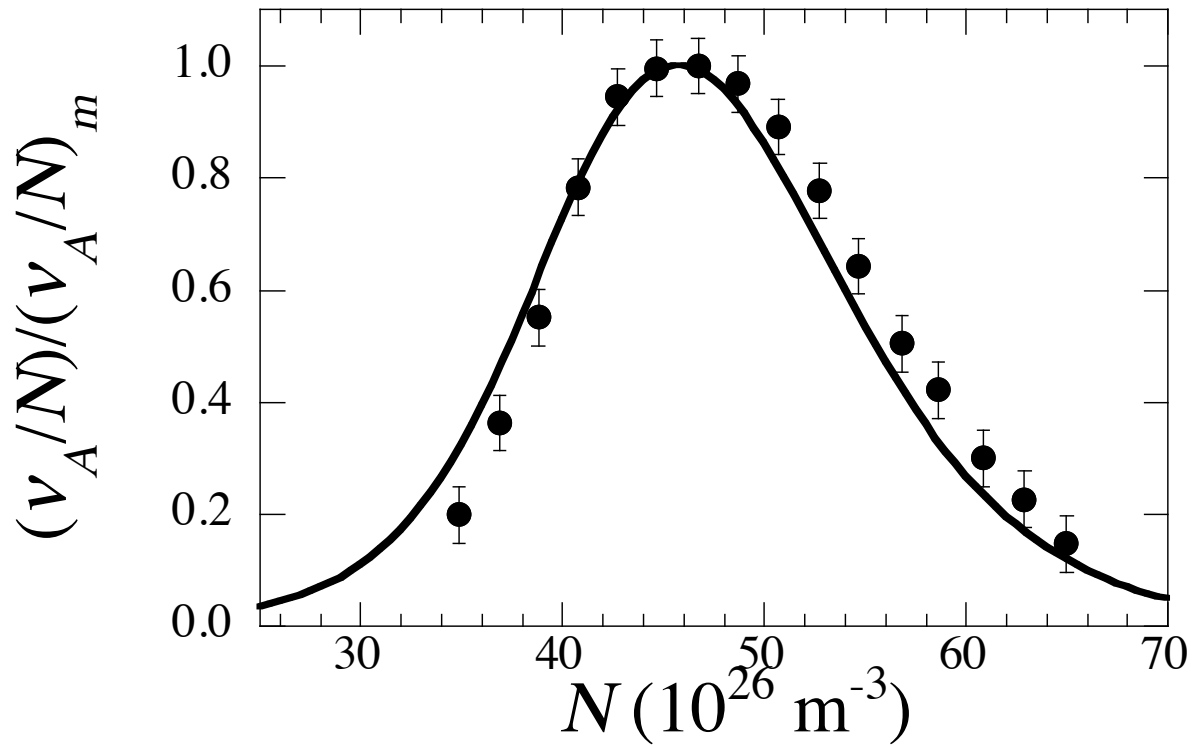}\caption{\small Normalized \(\nu_{A}/N\) vs \(N\) for \(T=80.1\,\)K. Solid line: \(c=0.225\).\label{fig:EggLTT80Kc0225}}
\end{figure}

At higher temperatures the agreement with the data is even better. For instance, in \figref{fig:EggLTT80Kc0225} we show the results for \(T=80.1\,\)K.
Similar results are obtained for all isotherms for a proper choice of the value of the parameter \(c\). 

\begin{figure}[t!]\centering\includegraphics[width=\textwidth]{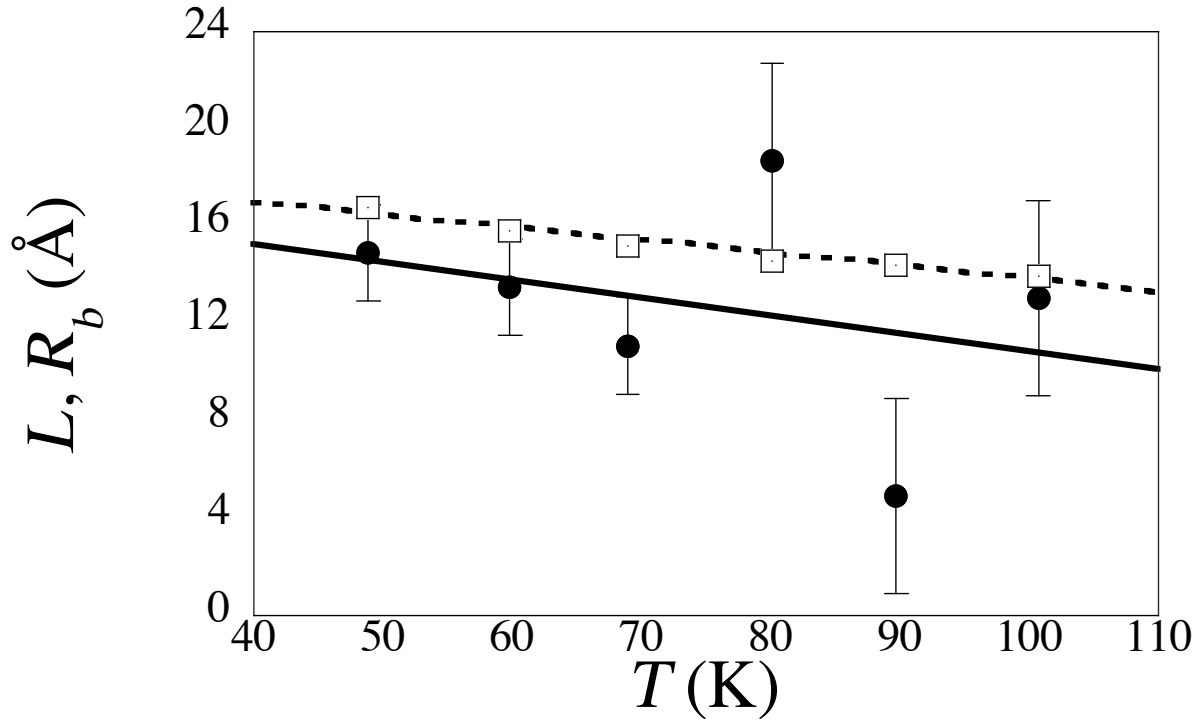}\caption{\small \(L(T)\) (closed symbols) and \(R_{b}\) (open symbols) vs \(T.\) The lines are only a linear fit to the data.\label{fig:LsamplingRbollavsT}}
\end{figure}

As a matter of fact, \(c\) and \(L\) turn out to be temperature dependent.  In \figref{fig:LsamplingRbollavsT} we compare the  values of \(L\) that give the best agreement with the peak shapes with the optimum ionic bubble radius \(R_{b}\).
In spite of the very large uncertainty in the sampling length especially at higher \(T\), we nonetheless note that \(L\) and \(R_{b}\) almost have the same value and temperature dependence. 
Probably, the attachment process more effectively  proceeds 
if the electron wavelength is close to that of the electron bound in the ion, which is of the order of \(R_{b}\). This may happen when a negative energy fluctuation of size \(L\) occurs around the molecule. 

This point of view could explain why \(L\) is larger for the 2nd peak than for the 1st one at the lowest temperature. Actually, the amplitude of vibration of the \(v^{\prime}=5\) level is larger than that of the \(v^{\prime}=4\) one, and,  presumably, the size of an actual ionic bubble should increase with  the  vibrational state of the ion.

\section{Conclusions\label{sect:conclusions}}
We have shown that the phenomenon of resonant electron attachment to O\(_{2}\) molecular impurities in a dense gas gives important pieces of information about the energetics and statistics of electron states in a disordered medium. 
Actually, the attachment frequency measurements sample the electron energy distribution function at the energy of resonance that is constant.

These new measurements in neon gas in broad density and temperature ranges once more confirm that the ground state energy of quasifree electrons in thermal equilibrium in a dense medium is offset by a density dependent contribution \(V_{0}\) that is the result of multiple scattering effects and that shifts the equilibrium distribution function. This shift affects many properties of quasifree electrons, including mobility~\cite{borghesani2002} and excimer formation~\cite{Borghesani2001}. The attachment process provides the researchers a tool to infer \(V_{0}\) without the shortcomings of other techniques~\cite{bruschi1984}.

At the same time, we have shown that the attachment process is very sensitive to the presence of fluctuations in the system. The shape of the attachment peak is related to the DOS of electrons in a disordered system.  
Actually, in neon the measurements are carried out at temperatures closer to the critical temperature than in helium. Thus, the fluctuations are stronger and their effect can be better put into evidence.  

In order to correctly describe the attachment process it is necessary to take into account the fact that the ion distorts the surrounding compliant medium giving origin to a complex structure that depends on the quantum nature of the outer electron in the ion and on its interaction with the electronic clouds of the host atoms. Only if this structure is accounted for a good agreement between experiment and theory can be achieved.

Finally, we would like to point out that the observation of an attachment peak from which so many pieces of information can be gathered can only happen if the electron-host atom interaction is dominated by short-range repulsive forces that lead to a positive \(V_{0}\). For negative scattering length gases as, for instance, argon in which the long-range polarization interaction is dominant, \(V_{0}\) is negative. In this case the resonance condition cannot be met and no peaks are observed although attachment still takes place. 
Moreover, 
 attaching molecular impurities other than O\(_{2}\) of different vibrational structure such as, for instance, SF\(_{6}\) could be exploited to sample the distribution function at different energies.

\appendix
\section{Excess free energy computation in the ionic bubble model\label{sect:deltF}}

The excess free energy of an ion surrounded by a cavity with respect to the free ion is reduced by the increase of the electron binding energy in the ion and is increased by the  work done against volume and surfaces forces to expand the cavity and form an interface
\begin{equation}
\Delta F = -\Delta \epsilon(R,T,N) +\frac{4\pi}{3}P(N,T)R^{3}+4\pi\sigma_{s}R^{2}+\epsilon_{P}
\label{eq:}\end{equation}
\(R\) is the cavity radius whose
optimum value \(R_{b}\) is obtained by minimizing \(\Delta F\) with respect to \(R\), thereby yielding \(\Delta F_{m}\) as a function of \(T\) and \(N\).
\(\Delta \epsilon (R,T,N)=\left[ \epsilon(R,T,N)-\epsilon_{A}\right]\) is the 
change of the electron binding energy in the ion and
\(\epsilon(R,T,N)\) is the lowest electron energy eigenvalue
in the field of the ion.  \(P(N,T)\) is the gas pressure given by the equation of state~\cite{katti1986}, and \(\sigma_{s}\) is the surface tension. \(\epsilon_{P}\) is the solvation energy of the ion immersed in the medium, which, according to~\cite{Miyakawa1969}, can be estimated by the Born solvation energy
 \begin{equation}
 \epsilon_{P} = - \frac{1}{2} \frac{e^{2}}{4\pi\epsilon_{0}R}\left(\frac{K-1}{K}\right)
 \label{eq:borgsolven}\end{equation} \(K\) is the relative dielectric constant of neon and is obtained, as usual, from the neon atomic polarizability \(\alpha_\mathrm{Ne}=2.66\,a_0^{3}\) via the Clausius-Mossotti equation \((K-1)/(K+2)=(4\pi/3) N\alpha_\mathrm{Ne}\). \(a_{0}\) is
 the Bohr radius. 
 
The \(s\)-wave eigenvalue is obtained for all \(T,\) \(N,\) and \(R\) by numerically solving the Schr\H odinger equation for the electron in a rectangular well 
subjected to the potential
\begin{equation}%
\label{cases}
V(r)=\cases{\infty& for $r \le R_{0}$\\
-\frac{1}{2}\frac{\alpha e^{2}}{4\pi\epsilon_{0}r^{4}}
&for $R_{0}<r\le R$\\
V_{0}(N) -\frac{1}{2}\frac{\alpha e^{2}}{4\pi\epsilon_{0}r^{4}}&for $r> R$}
\end{equation}
in which \(\alpha=10.6 a_{0}^{3}\) is the polarizability of O\(_{2}\).
\(R_{0}\) is the ion hard-core radius which is determined by solving the Schr\H odinger equation for the isolated ion with the constraint that the energy eigenvalue equals \(\epsilon_{A}\). The value \(R_{0}=0.909a_{0}\) is obtained that is very close to the value \(R_{0}=0.92	a_{0}\) obtained in literature by using the variational method~\cite{Khrapak1995a,volykhin1995}. By so doing, we implicitly neglect electrostriction and the distribution of host atoms outside the cavity.

As far as the surface term is concerned, 
an estimate of the surface tension \(\sigma_{s}\) %, though questionable as the system is gaseous, 
can be obtained by the parachor equation~\cite{macleod1923,hirschfelder1964,escobedo1996} by assuming that the cavity is empty
\begin{equation}
\sigma_{s} =\left(\pi_{c}N\right)^{4}
\label{eq:pc1}\end{equation}
\(\pi_{c}\) is the parachor constant whose crude estimate can be obtained from the surface tension \(\sigma_{t}\approx 5.5\times 10^{-3}\,\)N/m~\cite{Somayajulu1988} and  density \(N_{t}=373.9\times 10^{26}\,\)m\(^{-3}\) at the triple point as \(\pi_{c}= \sigma_{t}^{1/4}/\rho_{t}\approx 7.3\times 10^{-30}\,\)N/m\(^{1/4}\)m\(^{3}\). In any case, although the concept of surface tension in a gas and the parachor approximation might be questionable, the contribution of the surface energy term to \(\Delta F\) is rather small. For instance, for \(N\approx 40\times 10^{26}\,\)m\(^{-3}\) and for a typical cavity radius \(R\approx 8\,\)\AA\ we get \(4\pi R^{2}\sigma_{s}\approx 36\,\mu\)eV.

 \(V_{0}(N)\) is computed in Ref.~\cite{borghesani1990}. For the present computational purposes it can extremely well be approximated up to the highest densities by a 3rd order polynomial 
\begin{equation}
V_{0}(N) =7.096\times 10^{-4}N+6.743\times 10^{-6}N^{2}-7.883\times 10^{-9}N^{3} 
\label{eq:v0npoly}\end{equation} 
if \(N\) is expressed in units of \(10^{26}\,\)m\(^{-3}\) and \(V_{0}\) is in eV.

\begin{figure}[t!]\centering\includegraphics[width=\textwidth]{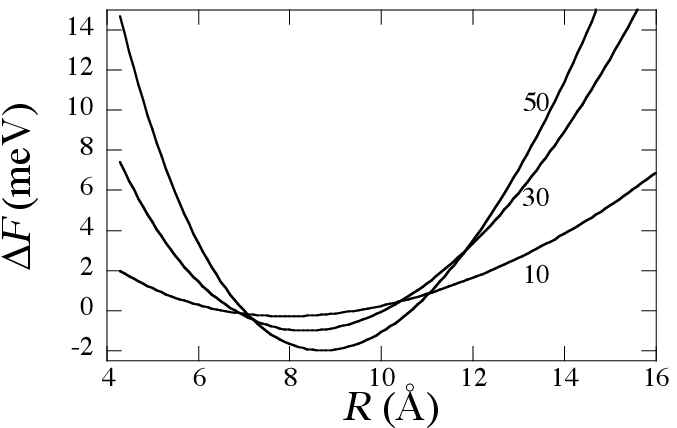}\caption{\small \(\Delta F\) vs \(R\) on some isopycnals for \(T=46.5\,\)K. The curves are labelled with the density value in units of \(10^{26}\)m\(^{-3}\). 
\label{fig:DeltaFvsRn103050T465}}
\end{figure}

\begin{figure}[t!]\centering\includegraphics[width=\textwidth]{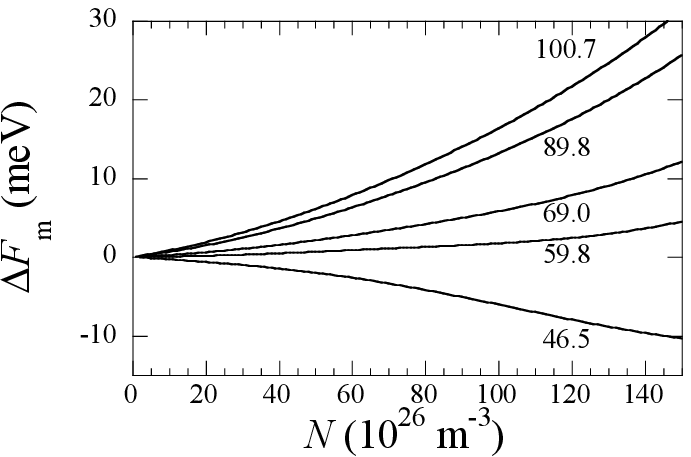}\caption{\small 
\(\Delta F_{m}\) vs \(N\) on the investigated isotherms. The curves are labelled with the value of \(T\).
\label{fig:deltaFminvsNT465to100}}
\end{figure}

 In \figref{fig:DeltaFvsRn103050T465} the results for \(\Delta F\) are shown as a function of \(R\) on some isopycnals for \(T=46.5\,\)K.
The curves are labelled by the density value in units of \(10^{26}\,\)m\(^{-3}\). All curves show a minimum as a function of the bubble radius \(R\) that gets deeper as the density is increased. The minimization procedure %of \(\Delta F\) 
yields \(\Delta F_{m}\)  and the optimum bubble radius \(R_{b}\) as a function of \(N\) for each isotherm. \(\Delta F_{m}	\) is plotted as a function of \(N\) for the investigated isotherms in \figref{fig:deltaFminvsNT465to100}.
By inspecting \figref{fig:deltaFminvsNT465to100} we note that, at constant \(N\),  \(\Delta F_{m}\)  increases with increasing \(T\) leading to an increase of the density \(N_{4}\) required by the resonance condition \eqnref{eq:rescondDF}. This increase is shown as the solid line in \figref{fig:N4THeNe} for neon.

\begin{figure}[b!]\centering\includegraphics[width=\textwidth]{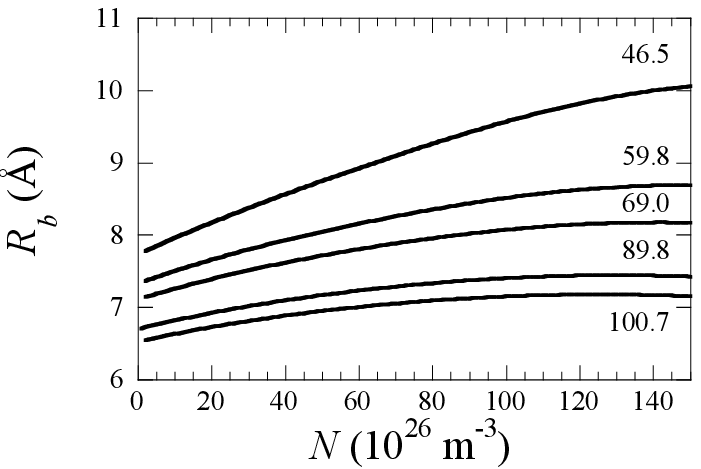}\caption{\small \(R_{b}\) vs \(N\). The curves are labelled by the temperature value.\label{fig:RbvsNT465to100}}
\end{figure}

As a byproduct of the computations also the optimum ionic bubble radius \(R_{b}\) is obtained. It is plotted as a function of \(N\) for the different isotherms in \figref{fig:RbvsNT465to100}.
The optimum radius \(R_{b}\) for a given \(N\) is larger at lower \(T\) mainly because it compensates for the lower pressure. Moreover, it has to be noticed that \(R_{b}\) shows a maximum at high \(N\) that shifts to lower density as \(T \) is increased. This behavior depends on both the density dependence of pressure and of the superlinear density dependence of \(V_{0}(N)\).

We finally note that this approach only gives the optimum state that minimizes the excess free energy and does neither give any pieces of information on the fluctuations of the free energy nor on the distribution of the bubble radii.

\clearpage
\section{The fluctuational model\label{app:egg}}
Here, we briefly describe the main features of the fluctuational model~\cite{Eggarter1970,Eggarter1971,Eggarter1972} with some results pertinent to the problem of attachment. The motion of quasifree electrons through a disordered medium of scatterers is converted to a motion in a smooth effective potential by
dividing the medium into boxes of side \(L\) and 
 averaging the scatterers distribution over them. The average potential is \(V_{0}(N)\) which fluctuates because the density is fluctuating from box to box within the volume of size \(L^{3}.\) The sampling length \(L\) can be considered as the scale of the autocorrelation of the potential. In each box, the DOS is assumed to be that relative to a freely propagating particle with energy above the local value of \(V_{0}\). The total DOS is obtained by summing over all boxes. By assuming that the fluctuations are normally distributed if \(L\) is not too small, the variance of the potential is given by
\begin{equation}
\sigma_{V}^{2}(N) = \frac{N}{L^{3}} S(0)\left[\frac{\partial V_{0}(N)}{\partial N}\right]^{2}
\label{eq:sv}\end{equation}
and the ensemble averaged DOS is
\begin{equation}
g(\epsilon,N) =\frac{1}{2\pi^{2}}\left(\frac{2m}{\hbar^{2}}\right)^{3/2}\sigma_{V}^{1/2}\mathcal{H}(x)
\label{eq:g}\end{equation}
in which \(x=[\epsilon - V_{0}(N)]/\sigma_{V}\).
The function \(\mathcal{H}(x)\) is given by  \begin{equation}
\mathcal{H}(x)= (2\pi)^{-1/2}\int\limits_{0}^{\infty}
z^{1/2}\mathrm{e}^{-[(x-z)^{2}/2]}\,\mathrm{d}z
\label{eq:h}\end{equation}
 The resulting DOS shows a low energy tail of non propagating electron states that is due to those boxes in which the fluctuations lead to a lower-than-average density.

Once the DOS is known, the normalized electron energy distribution function \(\mathcal{F}\) is given by
\begin{equation}
\mathcal{F}(\epsilon,T,N) =g(\epsilon,N) \mathrm{e}^{-\beta \epsilon}/ Q(T,N)
\label{eq:f}\end{equation}
in which \(\beta=(k_\mathrm{B}T)^{-1}\), \(k_\mathrm{B}\) is the Boltzmann constant,  and \(Q(T,N)= \int_{0}^{\infty}g(\epsilon,N)\mathrm{e}^{-\beta \epsilon}\,\mathrm{d}\epsilon\) is the partition function.

\begin{figure}[t!]\centering\includegraphics[width=\textwidth]{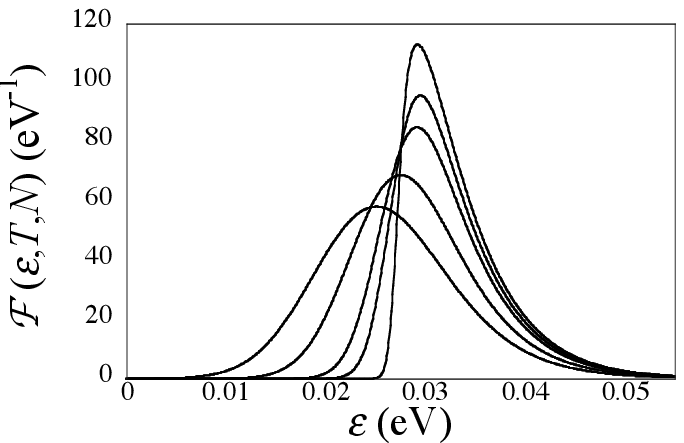}\caption{\small Normalized distribution function \(\mathcal{F}\) for \(T=48.4\,\)K and \(N=30\times 10^{26}\,\)m\(^{-3}.\) From left: \(c=0.2,\) 0.3, 0.4, 0.5, and 1. %The curves are labelled by the values of \(c\).
\label{fig:FdistT484N30c0251}}
\end{figure}

For the computations relative to attachment, the sampling length is chosen proportional to the electron thermal wavelength 
\begin{equation}
L= c(T) \frac{h}{\sqrt{2\pi m k_\mathrm{B}T}}
\label{eq:llt}\end{equation}
The effect of the choice of \(c\) can be observed by inspecting \figref{fig:FdistT484N30c0251} in which the normalized distribution function computed for \(T=48.4\,\)K and \(N=30\times 10^{26}\,\)m\(^{-3}\) is plotted for several values of \(c\).
First of all, we note that the distribution function is mainly shifted by \(V_{0}(N)\). In addition to that, we also notice that not only the distribution even more broadens as the value of \(c\) is decreased but also that its maximum shifts to lower energies. That is the reason why the attachment frequency peak position is determined in a combined way by both the shift due to \(V_{0}(N)\) and by that due to the fluctuations induced broadening of the distribution function.

\begin{figure}[t!]\centering\includegraphics[width=\textwidth]{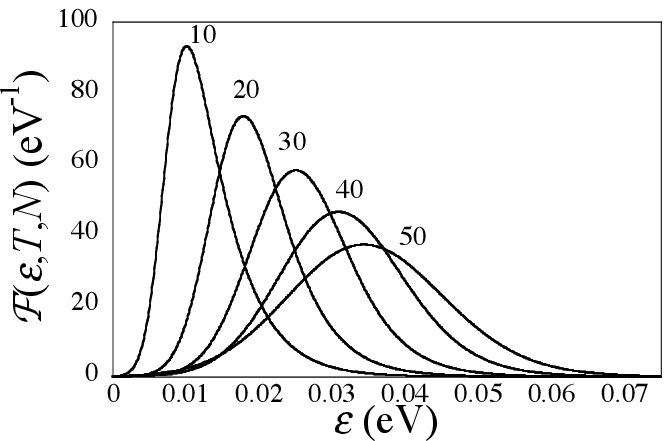}\caption{\small Normalized distribution function for \(T=48.4\,\)K and \(c=0.25\) for several \(N.\) The curves are labelled by the density values in units of \(10^{26}\,\)m\(^{-3}.\)\label{fig:FdistT484c025N10a50}}
\end{figure}

The effect of the density dependent shift \(V_{0}(N)\) on \(\mathcal{F}\) at constant \(L\) can be seen in \figref{fig:FdistT484c025N10a50} in which \(\mathcal{F}\) is plotted for several densities at constant \(T\).
%Although the sampling length is constant,
The distribution function broadens by increasing \(N\) because of the increase of \(S(0)\) with increasing \(N\) below the critical density, thereby leading to a larger effective potential variance.

\begin{figure}[h!]\centering\includegraphics[width=\textwidth]{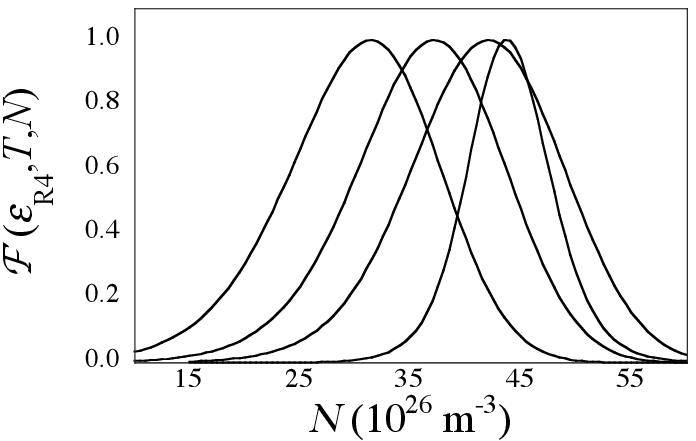}\caption{\small  \(\mathcal{F}(\epsilon_{R4},T,N)\) normalized to unity at the maximum for several values of \(c\) at \(T=48.4\,\)K. From left: \(c=0.125,\) 0.1375, 0.15, 0.3,\label{fig:ferT484Nseveralc}}
\end{figure}

Finally, the sensitivity of the normalized \(\nu_{A}/N\) on \(L\) can be realized by plotting the distribution function \(\mathcal{F}\) evaluated at the resonant energy \(\epsilon_{R4}\) as a function of \(N\) at constant \(T=48.4\,\)K for several values of the parameter \(c\) in \figref{fig:ferT484Nseveralc}.
By suitably choosing \(c\) one can control both the width of the resonance curve as well as the location of the maximum. In this way we also demonstrate that density dependent shift \(V_{0}(N)\) of the electron ground state energy is not sufficient to explain the features of the attachment peak because of the action the fluctuations 
are exerting on the distribution function via the density of states.

\clearpage
\section*{Acknowledgements}

The author gratefully acknowledges useful discussions with prof. M. Santini and also thanks prof. A. G. Khrapak for the critical reading of the manuscript.

\section*{References}
%\bibliography{MyCollection.bib}
\providecommand{\newblock}{}

\end{document}